\documentclass[fleqn,usenatbib]{mnras}

\usepackage{newtxtext,newtxmath}

\usepackage[T1]{fontenc}
\usepackage{ae,aecompl}

\usepackage{graphicx}	
\usepackage{amsmath}	
\usepackage{amssymb}	

\usepackage{graphicx}	
\usepackage{amsmath}	
\usepackage{amssymb}	

\newcommand{\Hz}{$\rm H(z)$~}
\newcommand{\fsig}{$f\sigma_8$(z)~}
\newcommand{\Hzfsig}{$\rm H(z)/H_0$-$f\sigma_8$(z)~}

\title[Constraints from cosmic growth and expansion]{Cosmological constraints from a joint analysis of cosmic growth and expansion}

\author[M. Moresco \& F. Marulli]{
M. Moresco$^{1,2}$\thanks{E-mail: michele.moresco@unibo.it}
and F. Marulli$^{1,2,3}$\thanks{E-mail: federico.marulli3@unibo.it}
\\
$^{1}$Dipartimento di Fisica e Astronomia, Universit\`a di Bologna, Via Gobetti 93/2, I-40129, Bologna, Italy\\
$^{2}$INAF - Osservatorio Astronomico di Bologna, Via Gobetti 93/3, I-40129, Bologna, Italy\\
$^{3}$INFN - Sezione di Bologna, Viale Berti Pichat 6/2, I-40127 Bologna, Italy
}

\date{Accepted XXX. Received YYY; in original form ZZZ}

\pubyear{2017}

\begin{document}
\label{firstpage}
\pagerange{\pageref{firstpage}--\pageref{lastpage}}
\maketitle

\begin{abstract}
Combining measurements on the expansion history of the Universe and on the growth rate of cosmic structures is key to discriminate between alternative cosmological frameworks and to test gravity. Recently, \cite{linder2017} proposed a new diagram to investigate the joint evolutionary track of these two quantities. In this letter, we collect the most recent cosmic growth and expansion rate datasets to provide the state-of-the-art observational constraints on this diagram. By performing a joint statistical analysis of both probes, we test the standard $\Lambda$CDM model, confirming a mild tension between cosmic microwave background predictions from Planck mission and cosmic growth measurements at low redshift ($z<2$). Then we test alternative models allowing the variation of one single cosmological parameter at a time. In particular, we find a larger growth index than the one predicted by general relativity $\gamma=0.65^{+0.05}_{-0.04}$). However, also a standard model with total neutrino mass of $0.26\pm0.10$ eV provides a similarly accurate description of the current data. By simulating an additional dataset consistent with next-generation dark-energy mission forecasts, we show that growth rate constraints at $z>1$ will be crucial to discriminate between alternative models.
\end{abstract}




\section{Introduction}

Since the discovery of the accelerated expansion of the
Universe \citep{riess1998,perlmutter1999}, different cosmological probes have been exploited to constrain the expansion history of the Universe and the growth rate of cosmic structures therein \citep[for a comprehensive review, see e.g.][]{weinberg2013}. The main quantities to be measured are the Hubble parameter, $\rm H(z)=$ $\dot{a}/a$, that describes the background expansion, and the linear growth rate $f$(z), defined as $f = d\ln G/d \ln a$, where $a$ is the scale factor, and $G$ is the growth factor of the matter density contrast. Usually, the quantity that is actually constrained is $f\sigma_8(z)$, where $\sigma_8$ is the matter power spectrum normalisation
at $8\, h^{-1}$Mpc.

Typically, \Hz and \fsig are measured separately using different cosmic probes, whose intrinsic properties make them more sensitive to some parameters, and less to others. For instance, type IA supernovae (SNe) trace luminosity distances up to $\rm z\sim1-1.5$, cosmic chronometers provide a direct measurement of \Hz up to $z\sim2$, redshift-space distortions constrain \fsig and baryon acoustic oscillations (BAO) give information on both \Hz and \fsig, but with less redshift coverage than the other probes. It is a common practice to combine different probes to increase the accuracy on the determination of 
cosmological parameters, but usually information on the growth factor and expansion are used disjointly \citep[but see e.g.][]{rapetti2013}. Recently, \cite{linder2017} proposed a new diagram exploiting these two quantities together. Specifically, it has been shown that in the $\rm H(z)/H_0$ vs. \fsig plane different cosmologies can be more easily disentangled.

In this letter, we take advantage of the most recent measurements of both \Hz and \fsig to explore the approach suggested by \citet{linder2017} from an observational perspective. The goal of this work is to collect the most recent observational data to provide the best available constraints on the \Hzfsig diagram. We used data from cosmic chronometers and redshift-space distortions to constrain possible extensions to the standard flat $\rm\Lambda CDM$ model and provide forecasts for next-generation galaxy redshift surveys.

\begin{figure*}
\includegraphics[width=\textwidth]{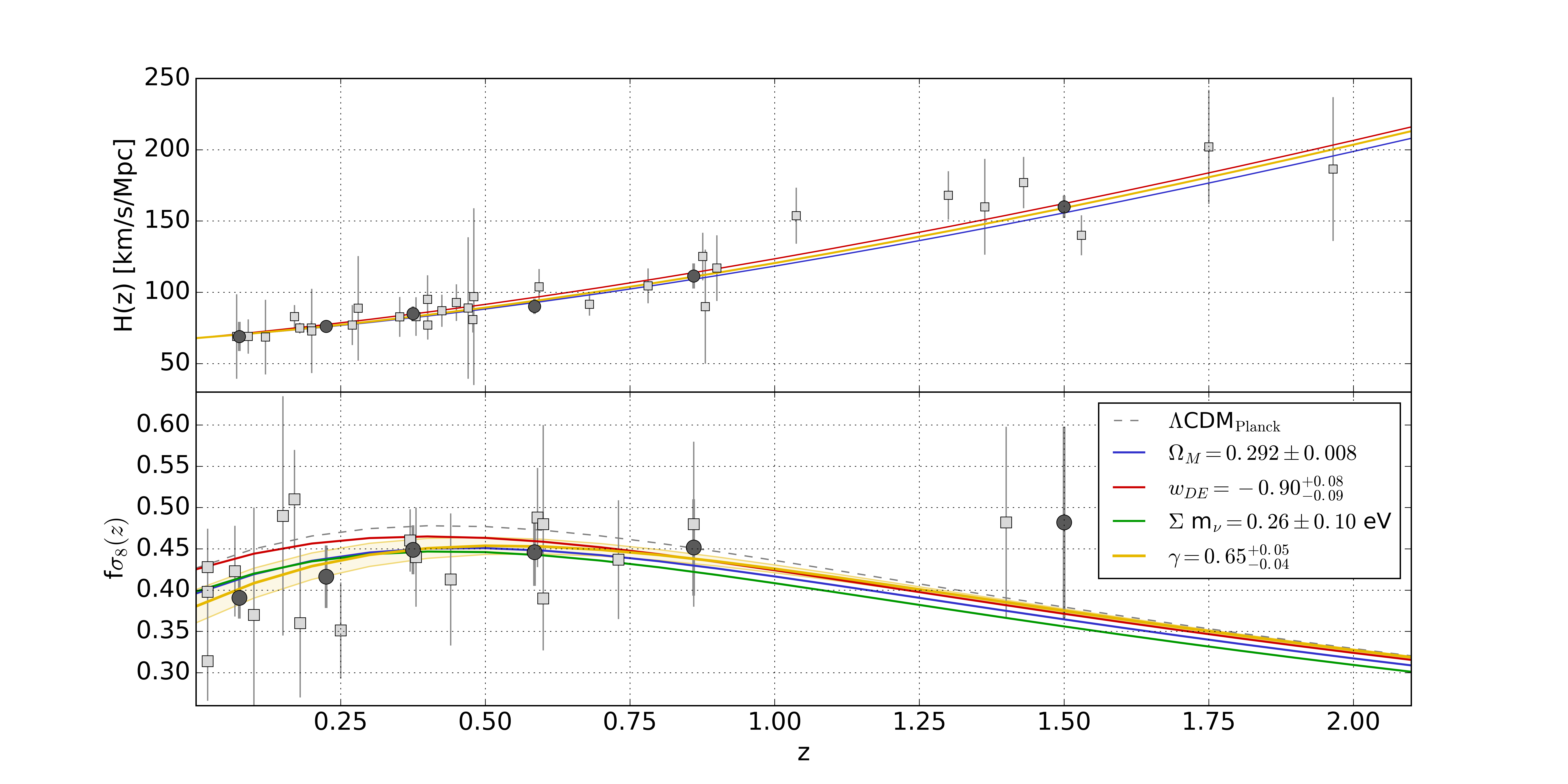}
\caption{The redshift evolution of the Hubble parameter, \Hz, (upper panel), and of the linear growth rate, \fsig (lower panel). The grey squares show the data used in this analysis, as described in Section \ref{sec1}. The black points show the binned data used to construct the \Hzfsig diagram. Best-fit models to \Hz and \fsig combined are shown with different lines: the dashed grey lines show the reference Planck2016 flat $\Lambda$CDM cosmology, while the coloured ones its extension, with free $\gamma$ (gold), $\Omega_M$ (blue), $\rm \Sigma m_{\nu}$ (green) and $w_{DE}$ (red). The yellow shaded areas show the $68\%$ confidence levels of the free $\gamma$ model, for illustrative purposes.}
\label{fig:fig1}
\end{figure*}


\section{Methods and data}
\label{sec1}
To construct the \Hzfsig diagram \citep{linder2017}, we collect the largest homogeneous dataset of \Hz and \fsig measurements, aimed at minimising any possible inconsistencies between different probes.

Differently from previous analyses \citep[e.g.][]{rapetti2013} that constrained the expansion history of the Universe based on indirect measurements, such as the luminosity distance $\rm D_L$(z) from SNe or the acoustic-scale distance ratio $\rm D_V$(z)/$r_d$ from BAO, here for the first time we rely only on direct constraints on the Hubble parameter \Hz obtained with the {\it cosmic chronometer} method. Originally proposed by \citet{jimenez2002}, this technique has been 
widely tested on different galaxy-redshift surveys, providing a direct estimate of \Hz without any cosmological assumption, over a large redshift range \citep[$0<z<2$, see][for a detailed discussion]{moresco2016}. In this work, we use in particular the measurements provided by \citet{simon2005, stern2010, moresco2012, zhang2014, moresco2015, moresco2016, ratsimbazafy2017}, that are reported in the upper panel of Fig.~\ref{fig:fig1}. We note that the cosmic chronometer method is quite new in the panorama of cosmological probes, and hence, while promising, it has not had the time yet to be studied to the same extent of more standard probes, such as BAO and SNe. We refer to \cite{weinberg2013} for a detailed review and comparison of the strengths and weaknesses of the various probes \citep[see also][for additional discussions]{guidi2015,liu2016,goddard2017}.

For the linear growth rate, we consider the \fsig dataset recently suggested by \cite{nesseris2017}, which collects only the independent measurements provided by \citet{percival2004, davis2011, hudson2012, turnbull2012, beutler2012, samushia2012, blake2012, blake2013, feix2015, howlett2015, huterer2016, chuang2016, okumura2016, delatorre2016}. These data are shown in the lower panel of Fig.~\ref{fig:fig1}. 

We analysed both datasets with a standard $\chi^2$ 
minimisation approach. As discussed in 
\citet{moresco2016} and \citet{nesseris2017}, the 
covariance matrix is diagonal for almost all 
measurements considered, except for the WiggleZ \fsig data, for which we used the covariance matrix provided by \citet{blake2012}.

As reference model, we consider the baseline flat $\Lambda$CDM model obtained by \citet{planck2016} (hereafter Planck2016), which assumes two massless and one massive neutrino with
mass 0.06 eV, $H_0=67.8$ km/s/Mpc, $\Omega_M=0.308$, $w_{DE}=-1$. We also set the value of the cosmic growth index $\gamma$ to $0.545$, as predicted by general relativity, where $f(z)\simeq\Omega_m(z)^\gamma$. As already discussed in previous works \citep[e.g.][]{macaulay2013, gilmarin2017, nesseris2017, marulli2017}, Planck2016 constraints are in some tension with low-redshift measurements, in particular with $f\sigma_8(z)$ constraints from recent redshift-space distortion analyses. This finding is confirmed also by the present work, as can be noted in the bottom panel of Fig.~\ref{fig:fig1}, that shows that Planck predictions overestimate, on average, \fsig measurements at $z<1$.

We explore four possible extensions to the reference $\Lambda$CDM model in order to get a better fit to the data, by changing each time one single parameter. Specifically, we vary the cosmic growth index $\gamma$, the matter density parameter $\Omega_M$, the total neutrino mass $\rm \Sigma m_{\nu}$, and the dark-energy equation of state parameter $w_{DE}$. The uncertainties in the current data are still too large to disentangle the degeneracies between the effects produced by some of these parameters, as will be shown in the following Section. Therefore, we decided to explore the effect of changing each parameter singularly.

We consider the following flat priors in the statistical analyses: $\gamma\in[0,1.5]$, 
$\Omega_M\in[0.1,0.6]$, $\rm \Sigma m_\nu\in[0,1.]$ eV, and $w\in[-2.,0.2]$. We note, however, that our results are not affected by the choice of these values, since all our results are well within the considered ranges.

To investigate the sensitivity of our data to the two probes, we perform both a fit separately to \Hz and \fsig, and a combined fit.
In order to construct the \Hzfsig diagram, we bin both datasets in 
the same redshift ranges chosen to get a uniform redshift sampling, having at least three points in both \Hz and $f\sigma_8$(z) bins, with the exception of the last two bins, where the sampling in \fsig is very scarce. In each redshift bin we estimate the variance weighted mean values of \Hz and \fsig. These data are reported in Fig.~\ref{fig:fig1} as a function of the mean redshift of the bins. This specific procedure is adopted purely for illustrative purposes (see Fig.~\ref{fig:fig2} and \ref{fig:fig3}), while all statistical analyses are performed on the original unbinned datasets.

To estimate \Hz and \fsig in the different cosmological models considered in this work, we exploit the \texttt{CosmoBolognaLib}, a large set of Open Source C++/Python libraries\footnote{Both the numerical libraries and the datasets analysed in this work are available at the following GitHub repository: \url{https://github.com/federicomarulli/CosmoBolognaLib} .}\citep{cosmobolognalib}.


\section{Analysis and discussion}
Fig.~\ref{fig:fig1} compares the best-fit models with the \Hz and \fsig datasets considered in this work. The values of the best-fit parameters of the four models considered, using \Hz and \fsig data both separately and combined together, are reported in Tab.~\ref{tab:tab1}. As previously stated, these results have been obtained by allowing the variation of one single parameter at a time. The goal is to quantify how the relaxation of each cosmological parameter can reduce the tension between the reference model and the data. The current measurement uncertainties are too large to allow the variation of more then one parameter. Indeed, as we verified, the constraints are approximately a factor of $5$ worse in the case of two parameters, making the analysis inconclusive.

\begin{figure}
	\includegraphics[width=0.5\textwidth]{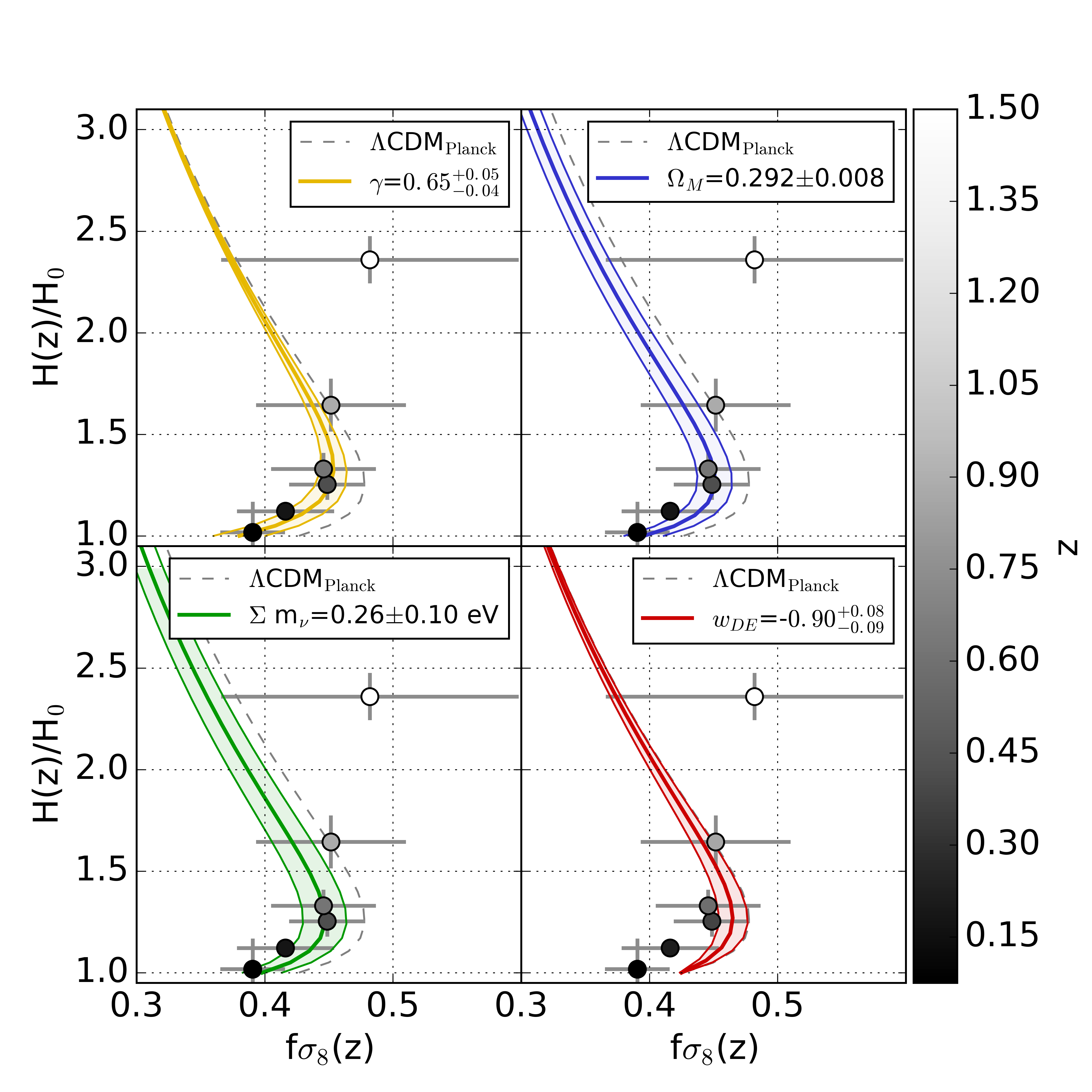}
    \caption{The \Hzfsig diagram \citep{linder2017}: cosmic expansion vs. cosmic growth. The four panels show the best-fit models and 1$\sigma$ associated errors (coloured lines and shaded area, respectively) obtained with the different extensions to the Planck2016 flat $\Lambda$CDM model, as indicated by the labels. The grey dashed lines show the reference Planck2016 cosmology. The points are the observed data binned in six redshift ranges, as in Fig.~\ref{fig:fig1}. Their colours indicate the mean redshift of each bin, as shown by the colour bar.}
    \label{fig:fig2}
\end{figure}

As expected, some parameters affect only the growth of structures (i.e. $\gamma$, and to a first approximation $\rm \Sigma m_{\nu}$), while the others have an impact also on the expansion history. In particular, all models show different evolutionary tracks that can be more clearly appreciated in the \Hzfsig diagram, shown in Fig.~\ref{fig:fig2}.
The data appear to be consistent with a higher value of $\gamma$ than the one predicted by general relativity, a lower value of $\Omega_M$ with respect to Planck2016 constraints, a value of 
$\rm \Sigma m_{\nu}$ significantly larger than the reference one, and a value of $w_{DE}$ smaller than the $\Lambda$ case. By simple relaxing one single parameter, it is possible to significantly reduce the tension between the data and the reference model. Both the $\gamma$, $\Omega_M$ and $\Sigma m_{\nu}$ models provide an accurate description of the data, in particular at low redshift ($z<0.5$). On the contrary, the $w_{DE}$ model does not provide an appreciably better fit, converging both at low and high redshifts to the Planck2016 reference model. 

As discussed above, we decided not to include in our datasets measurements of \Hz
from other probes, such as from BAO \citep{chuang2012,blake2012,fontribera2014,delubac2015}, to avoid mixing systematics from different probes that may bias the results. However, we verified that our results when including these data are consistent within 1$\sigma$ with the ones obtained with the dataset considered, except for $w_{DE}$, which results closer to -1 and yet more at odds with lower-redshifts \fsig measurements. We tested also different datasets of \fsig obtained with different techniques \citep[e.g.][]{pezzotta2016, hawken2016}, finding consistent results.

\begin{table*}
\centering
\caption{The best-fit values of the cosmological parameters let free to vary in the four models considered, using only \Hz data (first column), only \fsig data (second column), or using the two datasets combined together (third column). The fourth and fifth columns report, respectively, the values of $\Delta AIC_c$ and $\Delta BIC$ between the combined probes and the reference flat $\Lambda$CDM Planck2016 cosmology. The sixth and the seventh columns show the values of $\Delta AIC_c$ and $\Delta BIC$ when also the simulated data are included}.
\label{tab:tab1}
\begin{tabular}{lccccccc}
\hline
& H(z) & f$\sigma_{8}$(z) & combined & $\Delta AIC_c$ & $\Delta BIC$ & $\Delta AIC_c$ & $\Delta BIC$\\
& & & & \multicolumn{2}{c}{(combined vs. Planck2016)} & \multicolumn{2}{c}{(combined+simulated vs.Planck2016)}\\
\hline
$\gamma$ free & -- & $0.65^{+0.05}_{-0.04}$ & $0.65^{+0.05}_{-0.04}$ & 3.1 & 1.2 & 14.6 & 12.6\\
$\Omega_{\rm M}$ free & $0.33\pm0.03$ & $0.289\pm0.008$ & $0.292\pm0.008$ & 1.6 & -0.3 & 2.0 & 0.0\\
$\rm \Sigma m_{\nu}$ free & -- & $0.26\pm0.10$ & $0.26\pm0.10$ & 2.1 & 0.2 & 3.7 & 1.7\\
$w_{\rm DE}$ free & $-0.96^{+0.11}_{-0.12}$ & $-0.79^{+0.14}_{-0.15}$ & $-0.90^{+0.08}_{-0.09}$ & -0.5 & -2.4 & 4.3 & 2.3\\
\hline
\end{tabular}
\end{table*}

To test the significance of these results, we exploit two different selection model criteria, that is the Akaike Information Criterion \citep[hereafter AIC][]{akaike1974} and the Bayesian Information Criterion \citep[hereafter BIC][]{schwarz1978}. For the first criterion, we use the updated definition by \cite{sugiura1978}, which includes a correction when $N$ is small (hereafter $AIC_c$). These methods compare the best-fit likelihood functions of different models by weighting them by the number of free model parameters, thus penalising the overfitting of the data. The two criteria are defined as follows:
\begin{equation}
AIC_c=-2\ln \mathcal{L}_{max}+2k+\frac{2k(k+1)}{N-k-1}\,,
\end{equation}
\begin{equation}
BIC=-2\ln \mathcal{L}_{max} +k\ln N\,,
\end{equation}
where $\mathcal{L}_{max}$ is the maximum likelihood, $k$ is the number of the degrees of freedom of each model ($k=1$ in our cases), and $N$ is the number of data points. The BIC is the most conservative criterion between the two, disfavouring even more an increased number of free parameters. The 
differences between $AIC_c$ or $BIC$ values are then used to compare the models. Specifically, a model is considered to better represent the data on the base of the Jeffrey's scale \citep{jeffreys1961}. According to this scale, a difference smaller than 1 is {\em inconclusive}, between 1 and 2.5 is {\em moderate}, between 2.5 and 5 is {\em strong} and greater than 5 is {\em highly significant}. 
Compared to the reference model, we find that the data prefer a different value of $\gamma$ and $\rm \Sigma m_{\nu}$ with moderate to high significance (depending on the considered criterion), and of $\Omega_M$ with weak 
significance. On the other hand, the improvement that can be obtained with a different value of $w_{DE}$ turns out to be not significant. Indeed both the AICc and BIC indicate that the added $w_0$ parameter does not improve the fit in a statistically significant way with respect to the reference one.

To summarise, the data considered in this work suggest some deviations with respect to the flat $\Lambda$CDM model with one massive neutrino and Planck2016 cosmological parameters. We get a better match to the data by assuming a larger value of $\gamma$ with respect to the one predicted by general relativity. However, the current uncertainties in the data are too large to discriminate between this model and a standard model with massive neutrinos with $\rm \Sigma m_{\nu}\sim0.26$ eV. A similarly good agreement (but with a smaller confidence) can be obtained with a lower value of $\Omega_M$, though the required value would be in mild tension with Planck2016 constraints. Finally, changing the value of $w_{DE}$ has a marginal effect in the \Hzfsig diagram.

\begin{figure}
	\includegraphics[width=0.5\textwidth]{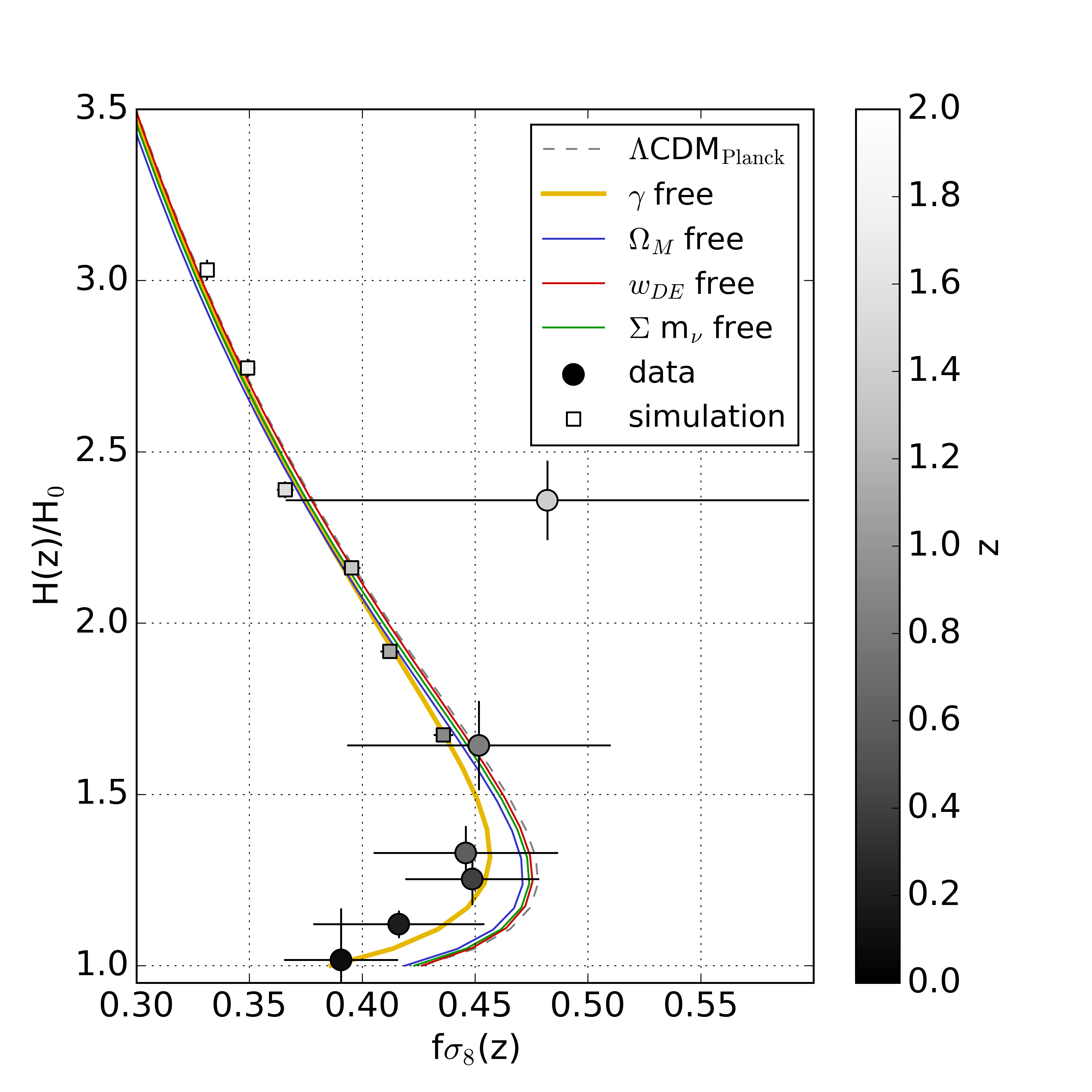}
    \caption{Same as Fig.~\ref{fig:fig2}, with the addition of simulated data (square points) forecasting constraints from next-generation galaxy redshift surveys. The simulated data assume a $\gamma=0.65$ extension to
$\Lambda$CDM. The best-fit models are shown with the same colour code as in Fig. \ref{fig:fig2}.}
    \label{fig:fig3}
\end{figure}

The \Hzfsig diagram appears particularly convenient to visualise the differences between alternative models, as can be appreciated in Fig. \ref{fig:fig2}. In particular, it can be noted that new measurements at $z\gtrsim1$ are required to
disentangle the effects of different parameters. To quantify this finding, we simulate some additional ($\rm H(z)$,$f\sigma_8$(z)) points at higher redshifts, that will be provided by future dark-energy missions, such as Euclid \citep{euclid} and WFIRST
\citep{wfirst}. In particular, we simulate six accurate ($\rm\sigma_{H(z)}/H(z)$=$\rm
\sigma_{f\sigma_8}$(z)$/f\sigma_8$(z)$=0.01$) measurements in the redshift range $0.9-2$, assuming as
the underlying model the $\gamma=0.65$ extension to
$\Lambda$CDM, that represents our best-fit to the current data. The assumed uncertainties are conservative, considered e.g. the available forecasts for the Euclid mission \citep{amendola2016}. Nevertheless, the goal of this test is just to provide rough estimates of the constraining power of next-generation galaxy redshift surveys, and it is not meant to be specifically designed to provide accurate forecasts for any specific future missions.

The new simulated data are presented in Fig.~\ref{fig:fig3}, together with the best-fit models discussed above. The differences with respect to the reference model are now extremely significant, as shown in Tab.~\ref{tab:tab1}. Moreover, being strongly constrained at high redshifts by these data, the models present now
significant deviations at low redshifts. In particular, the additional data would allow us to distinguish the effect of $\gamma$ and $\Sigma m_{\nu}$ at high significance, as reported in Tab.~\ref{tab:tab1}.

\section{Conclusions}
In this letter, we exploited the largest homogeneous dataset of \Hz and \fsig measurements currently available to construct the \Hzfsig diagram, recently introduced by \cite{linder2017}, testing the $\Lambda$CDM model and exploring possible extensions. We compared a reference flat $\Lambda$CDM model with four different extensions, each time varying a single cosmological parameter, namely $\gamma$, $\Omega_M$, $\rm \Sigma m_{\nu}$ and $w_{DE}$. We find that current low-redshift data appear in some tension with respect to the best-fit model obtained from the latest CMB analysis. Either a model with $\gamma=0.65^{+0.05}_{-0.04}$ or with $\Sigma m_{\nu}=0.26\pm0.10$ provides a better fit to the data at moderate to high statistical relevance, with respect to the reference model. Unfortunately, given the current measurement uncertainties, it is not possible to disentangle between these alternatives \citep{marulli2011}. We thus simulated six additional \Hz and \fsig measurements at $z\gtrsim1$ forecasting future dark-energy missions, such as Euclid and WFIRST, and found that these new 
data will allow us to distinguish between the models considered in this work with high statistical significance.

\section*{Acknowledgements}

MM and FM acknowledge the grants ASI n.I/023/12/0 ``Attivit\`a relative alla fase B2/C per la missione Euclid" and MIUR PRIN 2010-2011 ``The dark Universe and the 
cosmic evolution of baryons: from current surveys to Euclid" and PRIN MIUR 
2015 ``Cosmology and Fundamental Physics: illuminating the Dark Universe with Euclid".



\bibliographystyle{mnras}
\bibliography{bib} 






\bsp	
\label{lastpage}
\end{document}